\documentclass[aps,prx,twocolumn,showpacs,superscriptaddress]{revtex4-2}  

\usepackage{dirtytalk}
\usepackage{graphicx}  
\usepackage{dcolumn}   
\usepackage{bm}        
\usepackage{amsmath}
\usepackage{mathtools}
\usepackage{amsthm}
\usepackage{amssymb}
\usepackage{bbold} 
\usepackage{mathrsfs}
\usepackage{array}
\usepackage[dvipsnames,usenames]{color}
\usepackage{soul} 
\usepackage{ulem} 
\usepackage[bookmarks=true,colorlinks,linkcolor=blue,urlcolor=blue,citecolor=blue]{hyperref}

\newcommand{\be}{\begin{equation}}
\newcommand{\ee}{\end{equation}}
\newcommand{\bea}{\begin{eqnarray}}
\newcommand{\eea}{\end{eqnarray}}

\usepackage{physics}

\begin{document}

\normalem

\title{Topological Phase Transition in the Two-Leg Hubbard Model: Emergence of the Haldane Phase via Diagonal Hopping and Strong Interactions}

\author{João Pedro Gama D'Elia}
\affiliation{Instituto de Física, Universidade Federal do Rio de Janeiro, Rio de Janeiro, RJ, 21941-972, Brazil}

\author{Thereza Paiva} 
\affiliation{Instituto de Física, Universidade Federal do Rio de Janeiro, Rio de Janeiro, RJ, 21941-972, Brazil}

\begin{abstract}
We investigate the two-leg Hubbard model with diagonal hopping to explore the interplay between geometrical frustration and strong electron-electron interactions. Using the Density Matrix Renormalization Group (DMRG) method, we demonstrate the emergence of a topological Haldane phase, which results explicitly from the complementary effects of diagonal hopping-induced frustration and strong on-site Coulomb repulsion. The topological phase transition from a trivial insulator to the nontrivial Haldane phase is characterized by significant changes in magnetic properties, edge correlations, and the appearance of a nonzero string order parameter. Furthermore, we confirm the topological nature of this phase through a detailed analysis of the spin gap and entanglement spectrum, demonstrating clear signatures of symmetry-protected topological order.

\end{abstract}

\maketitle



\section{Introduction}


Geometrical frustration in condensed matter systems has emerged as a crucial concept for understanding a wide range of physical phenomena. When charge transport is introduced, the system's complexity increases due to electron hopping between nearest-neighbor and next-nearest-neighbor sites. The interplay between hopping and on-site interactions gives rise to nontrivial quantum phase transitions.

Hubbard ladder systems provide an important framework for exploring these phases. Their relevance has grown with the discovery of materials featuring coupled arrays of metal oxide ladders \cite{hiroi1991new,cava1991wf,norrestam1991new,johnston1987magnetic}. Moreover, they represent an intermediate regime between one-dimensional and two-dimensional systems, making them particularly well-suited for numerical studies. The ladder structure not only facilitates computational approaches but also enables the emergence of novel phases that are absent in purely one-dimensional chains.

Recent studies have examined Hubbard ladders in the context of high-temperature CuO$_2$ layer superconductors \cite{jiang2019superconductivity,zhou2023robust}. Band structure calculations have emphasized the significance of diagonal hopping in these systems \cite{andersen1995lda}, while angle-resolved photoemission spectroscopy \cite{shen1995electronic} and numerical studies have shown that diagonal hopping enhances superconductivity in Hubbard ladders \cite{jiang2019superconductivity}.

Traditionally, quantum phases in interacting systems follow the Landau-Ginzburg paradigm, where phases are characterized by local order parameters associated with spontaneous symmetry breaking. However, recent interest has shifted toward interaction-induced topological phases, particularly following the discovery of the fractional quantum Hall effect \cite{tsui1982two} and advances in quantum simulations with ultra-cold quantum gases \cite{cooper2019topological}, which allow precise control over interactions. Despite these advances, characterizing topological phases in interacting systems remains an ongoing challenge, as conventional topological invariants, such as the Chern number, rely on single-particle ground states.

A fundamental example of a symmetry-protected topological phase is the Haldane phase. Initially understood in the Affleck, Kennedy, Lieb, and Tasaki (AKLT) model, where it exhibited hidden antiferromagnetic order, given by a nonzero string order parameter, a non-vanishing gap in the thermodynamic limit \cite{affleck2004rigorous} and the presence of spin-1/2 edge states \cite{tasaki2020physics}. The exact solvability of the AKLT model, which possesses a valence bond solid (VBS) ground state, allowed for a deeper understanding of the Haldane phase as a phase protected by either $\mathbb{Z}_2\times\mathbb{Z}_2$, time-reversal or inversion symmetry \cite{tasaki2020physics}. The discovery of a gapped ground state in this spin-1 model helped to validate the Haldane conjecture further \cite{haldane1983continuum,haldane1983nonlinear} on integer-spin antiferromagnetic Heisenberg chains, leading to significant developments in the study of spin-1 Heisenberg models \cite{nightingale1986gap,parkinson1985spin} and the characterization of the Haldane phase.    

Previous works have explored the connection between frustrated spin ladders and spin chains, demonstrating that a fully frustrated spin-1/2 Heisenberg ladder can effectively realize a spin-1 Heisenberg chain in the Haldane phase \cite{honecker2000magnetization,honecker2016thermodynamic}. In fermionic systems, the Haldane phase has been shown to emerge under ferromagnetic exchange coupling between ladder rungs \cite{jazdzewska2023transition} and due to the unpairing of spin-1/2 edge sites \cite{anfuso2007string,moudgalya2015fragility}. Recent experiments have demonstrated the existence of the Haldane phase in a ladder system in an ultracold-atom quantum simulator \cite{sompet2022realizing}, showing the emergence of the Haldane phase even for a small system size (14 sites).

In this work, we demonstrate that the Haldane phase can arise due to a combination of geometrical frustration, introduced by diagonal hopping, and on-site Coulomb interactions. Our results reveal that the Haldane phase persists despite charge fluctuations and remains robust over a broad range of diagonal hopping amplitudes - contrary to expectations based on the associated Heisenberg ladders.

We organize our work as follows: in section \ref{sec:MM} we introduce the model and discuss the method used to solve it; we then discuss our results in different sections, starting with the strongly interacting (Heisenberg) limit, in section \ref{sec:HL}; magnetic properties in section \ref{sec:MP}; followed by results and discussions about the formation of edge states at \ref{sec:ES}, string order at \ref{sec:SO}, and the spin gap at 
\ref{sec:SG}. We study the entanglement spectrum in \ref{sec:ES}, and the charge gap in \ref{sec:CG}. Finally, we discuss our results at 
\ref{sec:D}.

\section{Model and Method}
\label{sec:MM}

\begin{figure}[h!]
    \centering    \includegraphics[width=\linewidth]{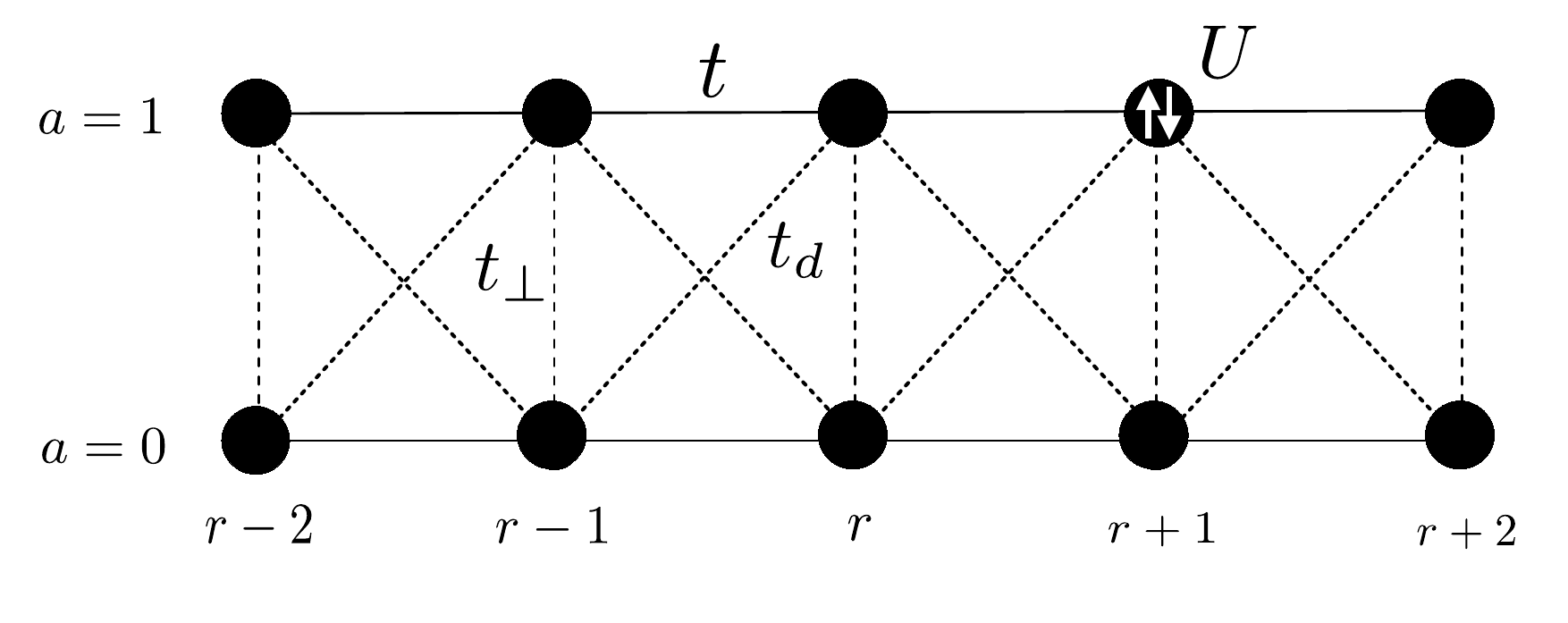}
    \caption{Schematic illustration of the two-leg Hubbard model, highlighting the hopping amplitudes between lattice sites: intra-leg hopping $t$ (solid lines), rung hopping $t_\perp$ (dashed lines), and diagonal hopping $t_d$ (dotted lines). The parameter $U$ is the Coulomb interaction strength that affects double-occupied sites. Rungs are labeled by $r \in \{0,\cdots,L-1\}$, where $L$ is the number of rungs and legs are labeled by $a \in \{0,1\}$.}
    \label{fig:system_illustration}
\end{figure}

The system consists of two identical Hubbard chains with nearest-neighbor hopping $t$, coupled by a perpendicular hopping $t_\perp$ and a next-nearest neighbor diagonal hopping $t_d$ (Figure \ref{fig:system_illustration}). Each leg (Hubbard chain) of the ladder is labeled by $a \in \{0,1\}$ and each rung (pair of sites connected by $t_\perp$) is labeled by $r \in \{0,\cdots,L-1\}$, where $L$ is the number of rungs.
The Hamiltonian that governs the system is given by 
\begin{equation}
\begin{gathered}
 \label{eqn:full_hamiltonian}
    H = -t\sum_{a,r,\sigma}(c_{a,r,\sigma}^{\dagger}c_{a,r+1,\sigma} + H.c.)\\
    - t_\perp\sum_{r,\sigma}(c_{0,r,\sigma}^{\dagger}c_{1,r,\sigma} + H.c.)\\
    - t_d\sum_{r,\sigma}(c_{0,r,\sigma}^{\dagger}c_{1,r+1,\sigma} + c_{1,r,\sigma}^{\dagger}c_{0,r+1,\sigma} + H.c.)\\
    + U\sum_{a,r} n_{a,r,\uparrow}n_{a,r,\downarrow}
\end{gathered}
\end{equation}
where $c_{a,r,\sigma}^{\dagger}$ ($c_{a,r,\sigma}$) is the creation (destruction) operator of a fermion on leg $a$ and rung $r$ with spin-1/2 projection in the $z$ direction $\sigma \in \{\uparrow,\downarrow\}$. The last term accounts for the on-site Coulomb repulsion $U$, which affects fermions occupying the same site. We set the hopping $t$ as our energy unit, and along this work, we set $t_\perp =t$, focusing our investigation in the roles of $t_d$ and $U$. All our results will be taken at half-filling.

We have used the Density Matrix Renormalization Group  (DMRG) method \cite{schollwock2011density,verstraete2023density} to numerically solve Hamiltonian \ref{eqn:full_hamiltonian}. In the DMRG method the many-body wavefunction is represented as a sequence of tensors. Each tensor encodes local physical degrees of freedom along with virtual indices that capture the entanglement between neighboring sites, and the overall state is constructed by contracting these tensors sequentially. The optimization process involves iteratively updating the tensors to minimize the energy expectation value while constraining the bond dimension, which controls the amount of entanglement retained.

We probed system sizes ranging from $L = 28$ rungs to $L = 112$. In most calculations, we have reached a number of density-matrix eigenstates up to $m = 2000$, conserving the total number of particles and the total magnetization $S_z$. The maximum truncation error ranged from $10^{-9}$ to $10^{-11}$. Our calculations were set to cover a wide range of parameters, with $t_d/t$ ranging from $0$, the non frustrated bipartite system, to $1$, and $U/t$ up to $16$.

\begin{figure}
    \centering
    \includegraphics[width=\linewidth]{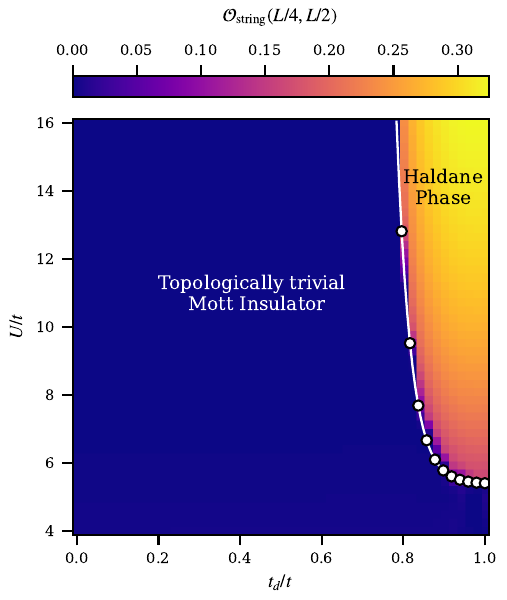}
    \caption{Phase diagram of the two-leg Hubbard model with diagonal hopping, determined from the numerical calculation of the string order expectation value (see equation \ref{eqn:string_order}). The color intensity indicates the magnitude of the string order parameter, distinguishing between the topologically trivial Mott insulating region (blue) and the nontrivial Haldane phase (yellow).}
    \label{fig:phase_diagram}
\end{figure}

Our main results are presented from the outset:  we demonstrate the existence of the Haldane phase, with a phase diagram in the  ($t_d$, $U$) plane, shown in Figure \ref{fig:phase_diagram}.  In the following sections, we explain how the determination and characterization of the Haldane phase was performed and how the critical line was obtained.

\section{Heisenberg limit}
\label{sec:HL}

In the strongly interacting limit: $U \rightarrow \infty$, a Schrieffer–Wolff unitary transformation \cite{bravyi2011schrieffer} can be applied to map the Hamiltonian (equation \ref{eqn:full_hamiltonian}) into the two-leg antiferromagnetic Heisenberg model
\begin{equation}
\begin{gathered}
    H = J\sum_{a,r}\mathbf{S}_{a,r}\cdot\mathbf{S}_{a,r+1} - J_\perp\sum_{r}\mathbf{S}_{0,r}\mathbf{S}_{1,r}\\
    + J_d\sum_{r}(\mathbf{S}_{0,r}\mathbf{S}_{1,r+1,\sigma} + \mathbf{S}_{1,r}\mathbf{S}_{0,r+1})
    \label{eqn:heisenberg}
\end{gathered}
\end{equation}
where the exchange couplings are given by $J_x \approx \frac{4t_x^{2}}{U}$. When $J_d = J$, the Hamiltonian depends only on the total spin of each rung $\mathbf{T}_r = \sum_{a=0}^{1} \mathbf{S}_{a,r}$ \cite{honecker2000magnetization} and can be rewritten as
\begin{equation}
     H = J\sum_{r}\mathbf{T}_{r}\cdot\mathbf{T}_{r+1} + J_\perp\sum_{r}\left(\frac{\mathbf{T}_r^{2}}{2} - \frac{3}{4}\right)
     \label{eqn:heisenberg_1}
\end{equation}
where $\mathbf{T}_r^{2} = T_r(T_r+1)$ and $T_r$ takes values in $\{0,1\}$.

For large $J_\perp$, the system favors a rung-singlet phase ($T_r = 0$ for all rungs), with ground state energy per rung $E_S = -3J_\perp/4$. In contrast, for small $J_\perp$ the system enters a rung-triplet phase ($T_r = 1$ for all rungs), effectively behaving as a spin-1 Heisenberg model in the Haldane phase, where each rung of the system is associated with a spin-1 particle. In this case, the ground-state energy per rung is given by \cite{honecker2016thermodynamic}
\begin{equation}
    E_T = J\cdot\epsilon_\text{H} + J_\perp/4
    \label{eqn:rung_triplet_energy}
\end{equation}
where $J\cdot\epsilon_\text{H} \approx -1.401484 J$\cite{white1993numerical,golinelli1994finite} is the ground state energy per site of the spin-1 Heisenberg model. The transition between these two distinct phases occurs at $J_{\perp c} \approx |J\cdot\epsilon_\text{H}|$ \cite{honecker2000magnetization}.

The possibility of achieving an effective spin-1 chain in a topologically non-trivial phase from a two-leg ladder motivates the study of the Hamiltonian (equation \ref{eqn:full_hamiltonian}) for finite values of $U$, where charge fluctuations are allowed and favored by the hopping terms, and for $t_d \neq t$, where the local conservation of each total rung spin $\mathbf{T}^{2}_r$ is no longer strictly enforced in the Heisenberg limit. The central question we explore is whether the Haldane phase persists under these conditions, where charge fluctuations are present and $t_d \ne t$.

We can further explore the result of equation \ref{eqn:rung_triplet_energy}. Let $\ket{\psi}$ be the ground state of Hamiltonian \ref{eqn:heisenberg_1} in the Haldane phase, we can express $\epsilon_\text{H}$ as 
\begin{equation}
\begin{aligned}
    \epsilon_\text{H} &= \frac{1}{L} \sum_r\bra{\psi}\mathbf{T}_r\mathbf{T}_{r+1}\ket{\psi}\\
    &= \frac{1}{L} \sum_{\beta \in \{x,y,z\}}\sum_r\bra{\psi}T_r^{\beta}T_{r+1}^{\beta}\ket{\psi}
    \label{eqn:e_h_beta}
\end{aligned}
\end{equation}
Since the Hamiltonian \ref{eqn:heisenberg_1} has SU(2) symmetry, we can use the spin rotation operators, defined as
\begin{equation}
    \mathcal{U}_\phi^{\beta}  \equiv \prod_{r=0}^{L-1}\exp{-i\phi T^{\beta}_r}
\end{equation}
to find that,
\begin{equation}
\begin{aligned}
    \bra{\psi} T_r^{x}T_{r+1}^{x}\ket{\psi} &= \bra{\psi}\mathcal{U}_{\frac{\pi}{2}}^{y\dagger}\left(\mathcal{U}_{\frac{\pi}{2}}^{y} T_r^{x} \mathcal{U}_{\frac{\pi}{2}}^{y\dagger}\right)\left(\mathcal{U}_{\frac{\pi}{2}}^{y} T_{r+1}^{x}\mathcal{U}_{\frac{\pi}{2}}^{y\dagger}\right)\mathcal{U}_{\frac{\pi}{2}}^{y} \ket{\psi}\\
    &= \bra{\psi} T_r^{z}T_{r+1}^{z}\ket{\psi}
\end{aligned}
\end{equation}
where the same argument can be used for the $\beta = y$ spin rung operator term in equation \ref{eqn:e_h_beta}, the only difference is that for that case we use the $\mathcal{U}_{\frac{-\pi}{2}}^{x\dagger}$ unitary operator. With that, we have
\begin{equation}
    \epsilon_\text{H} = \frac{3}{L} \sum_r\bra{\psi}T_r^{z}T_{r+1}^{z}\ket{\psi}
\end{equation}
Now, the dependence only on the total spin of each rung leads to an equivalence between leg and diagonal correlations: 
\begin{equation}
    \langle S^{z}_{a,r}S^{z}_{a^\prime,r+1}\rangle = \langle S^{z}_{a,r}S^{z}_{a,r+1}\rangle
    \label{eqn:sym_relation}
\end{equation}
such that using equation \ref{eqn:sym_relation} we can write
\begin{equation}
    \bra{\psi}T_r^{z}T_{r+1}^{z}\ket{\psi} = 4\bra{\psi} S_{0,r}^{z}S_{1,r+1}^{z}\ket{\psi}
\end{equation}
Finally, we can express $\epsilon_H$ as
\begin{equation}
    \epsilon_\text{H} = \frac{12}{L} \sum_r\bra{\psi}S_{0,r}^{z}S_{1,r+1}^{z}\ket{\psi}
\end{equation}
Translation invariance then implies
\begin{equation}
    \epsilon_\text{H} = 12\bra{\psi}S_{0,r}^{z}S_{1,r+1}^{z}\ket{\psi}
\end{equation}
Substituting $\epsilon_\text{H} = -1.401484$, we have
\begin{equation}
    12\bra{\psi}S_{0,r}^{z}S_{1,r+1}^{z}\ket{\psi} = -1.401484
    \label{eqn:diag_heis}
\end{equation}

\begin{figure}[h!]
    \centering
    \includegraphics[width=\linewidth]{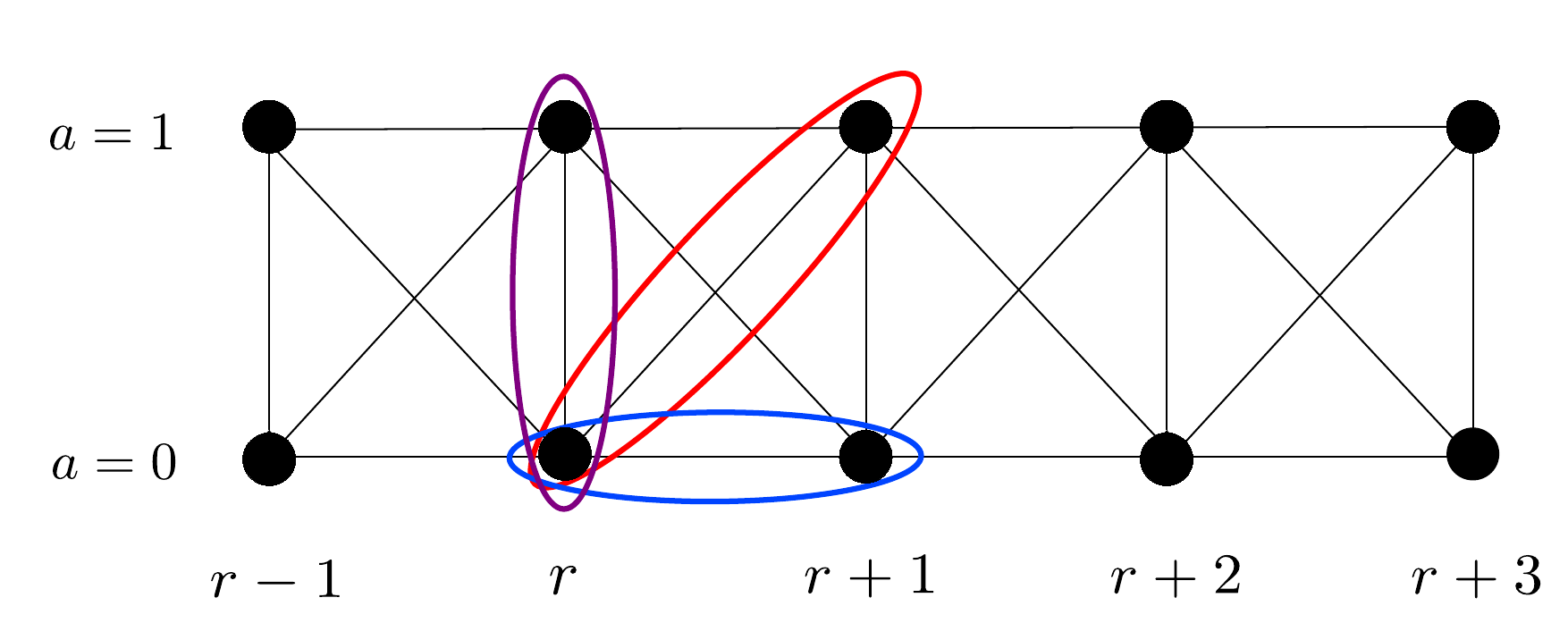}
    \caption{Schematic representation of spin-spin correlations on the two-leg ladder. The purple line corresponds to rung correlations $\langle \mathcal{O}_{0,r}\mathcal{O}_{1,r}\rangle$, the red line denotes diagonal correlations $\langle \mathcal{O}_{0,r}\mathcal{O}_{1,r+1}\rangle$, and the blue line illustrates leg correlations $\langle \mathcal{O}_{0,r}\mathcal{O}_{0,r+1}\rangle$.}
    \label{fig:neighbor_correlations}
\end{figure}

The above equation implies that in the Heisenberg model with $J_d = J$, spin-spin correlations between neighboring diagonal sites must always be negative in the Haldane phase. In contrast, in the non-frustrated (bipartite) limit ($t_d = 0$), the spin-spin correlation functions exhibit a characteristic sign structure: spin-spin correlations between sites on different sublattices are always negative, while those on the same sublattice remain positive \cite{white1993numerical}. This holds for all positive values of $U$, including the strong-coupling Heisenberg limit (equation \ref{eqn:heisenberg} with $J_d = 0$). In particular, in the $t_d = 0$ limit, for leg, rung and diagonal spin-spin correlation function on neighboring sites (illustrated in Figure \ref{fig:neighbor_correlations}), we have 
\begin{equation}
\begin{aligned}
    \langle S^{z}_{a,r}S^{z}_{a,r+1}\rangle &< 0 \text{ (leg)},\\
    \langle S^{z}_{0,r}S^{z}_{1,r}\rangle &< 0 \text{ (rung)},\\    
    \langle S^{z}_{a,r}S^{z}_{a^\prime,r+1}\rangle &> 0 \text{ (diagonal)}
\end{aligned}
\end{equation}
This means that as we increase $t_d/t$ from $0$ to $1$ and $U/t$ from $0$ to $\infty$, at some point the spin-spin correlation function between neighboring diagonal sites must change from positive to negative values.

\section{Magnetic Properties}
\label{sec:MP}

 As was pointed out in the previous section, the transition from a topologically trivial state to the Haldane phase on the two-leg Hubbard model with diagonal hopping must be accompanied by a change in the local magnetic properties of the system. In particular, the spin-spin correlation function between neighboring diagonal sites must change sign as $t_d/t \rightarrow 1$ and $U/t \rightarrow \infty$, and approach  $\langle S^{z}_{a,r}S^{z}_{a^\prime,r+1}\rangle = \epsilon_\text{H}/12$ in the Heisenberg limit in the Haldane phase.
 
\begin{figure} 
    \centering
    \includegraphics[width=0.9\linewidth]{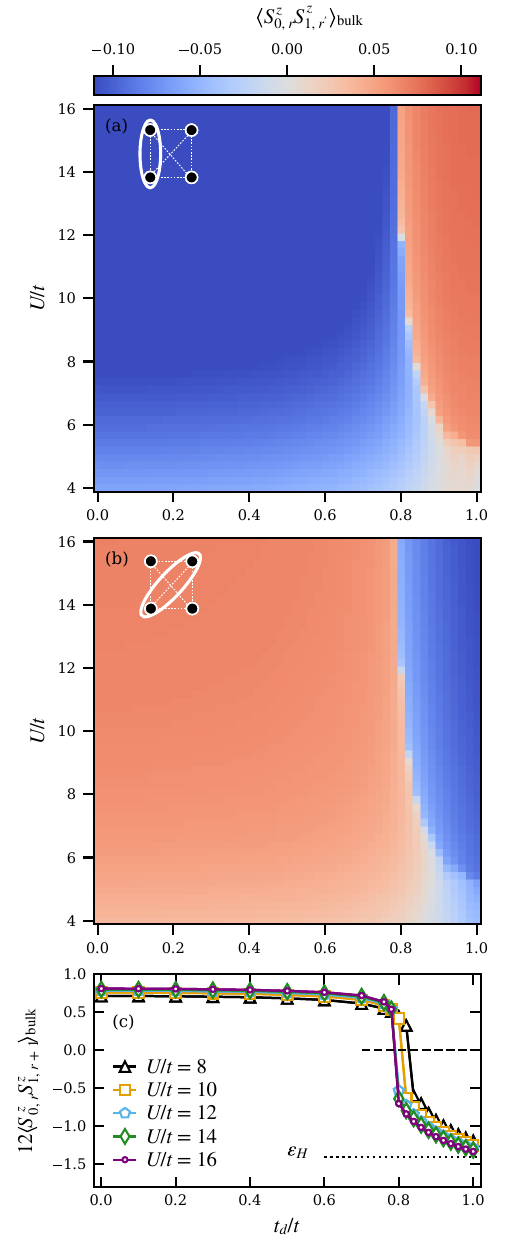}
    \caption{Average spin-spin correlation functions calculated in the bulk of the ladder, showing their dependence on diagonal hopping $t_d$ and interaction strength $U$ for a $L = 112$ rung system. Panels (a) and (b) present the rung and diagonal correlations, respectively. Panel (c) explicitly shows the evolution of diagonal correlations as $t_d$ increases for different values of $U$, highlighting the transition from positive to negative correlations indicative of the emergence of the Haldane phase. The dashed line marks zero correlation, and the dotted line indicates the theoretical limit derived from the spin-1 Heisenberg model. } 
    \label{fig:rung_diag_corr}
\end{figure}

 In Figures \ref{fig:rung_diag_corr}(a) and \ref{fig:rung_diag_corr}(b), we analyze the behavior of spin-spin correlations as a function of the diagonal hopping amplitude ($t_d/t$) and the interaction strength ($U/t$). Specifically, Figure \ref{fig:rung_diag_corr}(a) presents the average spin-spin corelation function between sites in the same rung $\langle S^{z}_{0,r}S^{z}_{1,r}\rangle_\text{bulk}$, whereas Figure \ref{fig:rung_diag_corr}(b) shows the spin-spin correlation function between neighboring diagonal sites $\langle S^{z}_{0,r}S^{z}_{1,r+1}\rangle_\text{bulk}$. Since we work with open-boundary conditions, both correlation functions were calculated in the system´s bulk, averaging over rungs from $L/4$ to $3L/4$, to minimize boundary effects
 \begin{equation}
     \langle S^{z}_{0,r}S^{z}_{1,r^\prime}\rangle_\text{bulk} = \frac{1}{L_\text{bulk}}\sum_{r = r_0}^{r_f} \langle S^{z}_{0,r}S^{z}_{1,r^\prime}\rangle
 \end{equation}
 with $r_0 = L/4$, $r_f = 3L/4$ and $L_\text{bulk} = r_f-r_0$.

\begin{figure}
    \centering
    \includegraphics[width=0.90\linewidth]{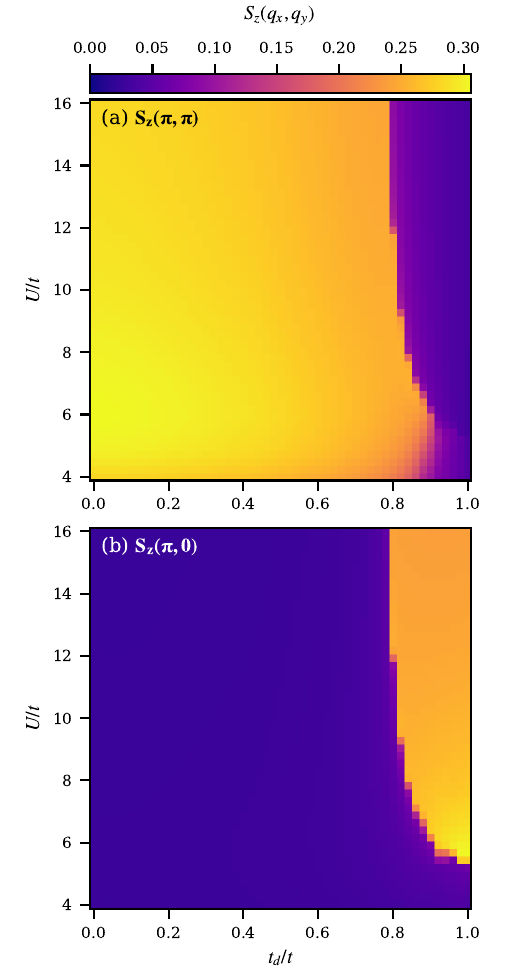}
    \caption{Spin structure factor $S_z(q_x,q_y)$ calculated at selected momentum points as functions of $t_d/t$ and $U/t$ for a $L=112$ rung system. Panel (a) shows the structure factor at $(q_x,q_y)=(\pi,\pi)$, reflecting antiferromagnetic correlations along both legs and rungs, whereas panel (b) corresponds to $(q_x,q_y)=(\pi,0)$, indicating antiferromagnetic correlations along legs and ferromagnetic correlations along rungs. The behavior of the structure factor differentiates the magnetic arrangements inside and outside the topological phase region. }
    \label{fig:ssf}
\end{figure}
\begin{figure}[]
    \centering
    \includegraphics[width=\linewidth]{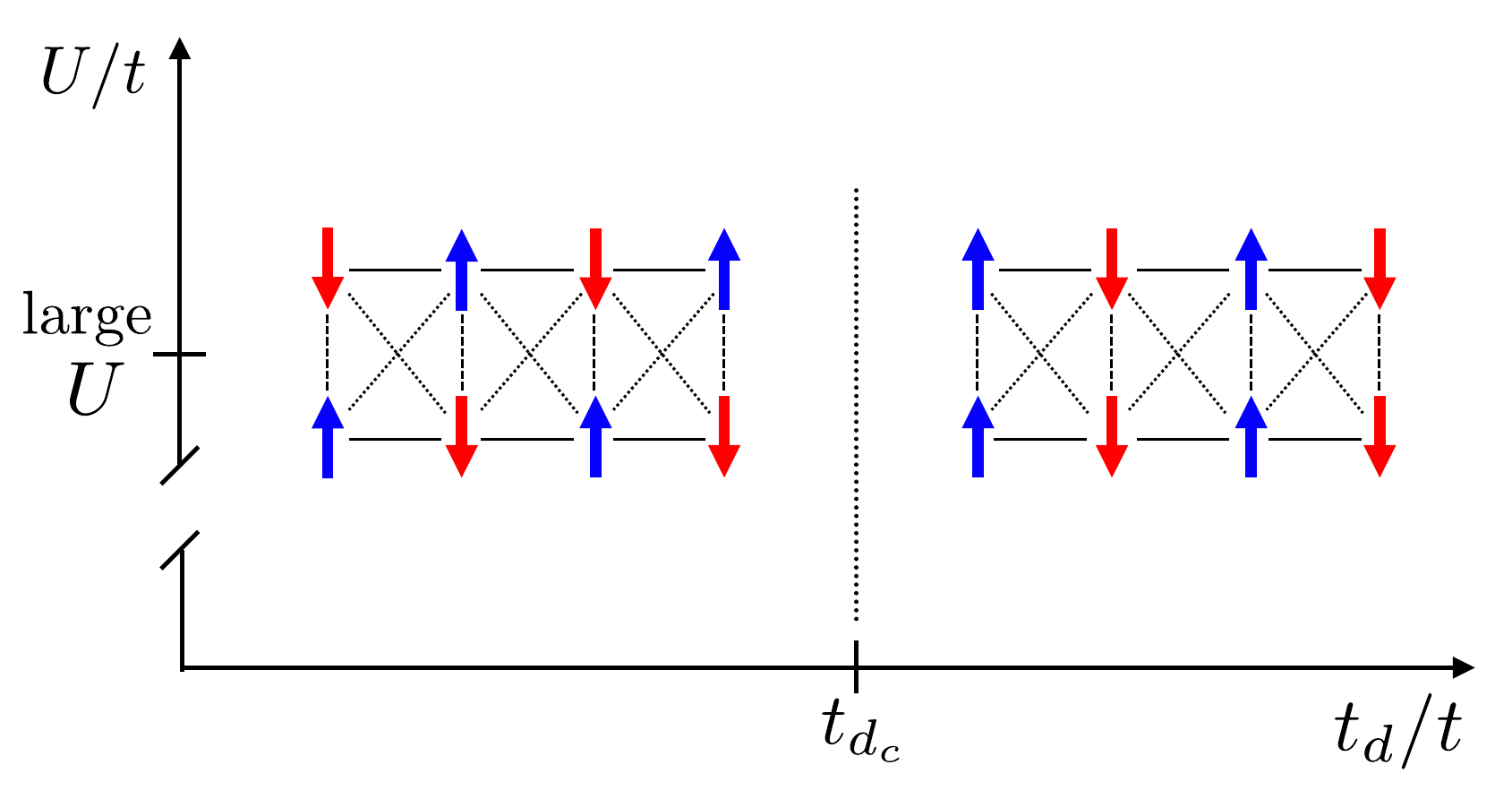}
    \caption{Illustrative depiction of distinct magnetic orderings identified in the two-leg Hubbard model in the strong-coupling regime: (left) rung-singlet dominated trivial phase and (right) rung-triplet.}
   \label{fig:mag_ordering}
\end{figure}

The results illustrate the expected sign change in diagonal spin-spin correlations in a curve dependent on $t_d$ and $U$ in the $t_d\times U$ plane, demarcating a dome. Figure \ref{fig:rung_diag_corr}(c) provides additional insight by depicting the evolution of diagonal spin-spin correlations as $t_d/t$ increases for several fixed values of $U/t$, demonstrating the correlation shift from positive to negative values. It also shows $t_d/t = 1$ points approaching the theoretical value from the spin-1 Heisenberg chain $\epsilon_\text{H}/12$ as U/t becomes large, in agreement with the theoretical analysis of the spin-1/2 Heisenberg ladder in the Haldane phase (see section \ref{sec:HL}). For the rung correlations, we also see a change in sign for points inside the dome; this may be associated with the predominance of rung triplets, as expected in the Haldane phase. Collectively, these observations confirm that the magnetic ordering changes distinctly across the identified dome region under the combined effect of diagonal hopping-induced frustration and strong Coulomb interactions.

Figures \ref{fig:ssf}(a) and \ref{fig:ssf}(b) explore the spin structure factor $ S_z(q_x,q_y)$, defined below, at different momentum points, providing complementary information on the system’s magnetic arrangement.

\begin{gather}
    \label{eqn:ssf}
    S_z(q_x,q_y) = \frac{1}{2L_\text{bulk}}\sum_{a,a^\prime}\sum_{r,r^\prime =r_0}^{r_f} e^{iq_x(r-r^\prime)}e^{iq_y(a-a^\prime)}\langle S^{z}_{a,r} S^{z}_{a^\prime,r^\prime} \rangle
\end{gather}
Figure \ref{fig:ssf}(a) examines $S_z(\pi,\pi)$, which characterizes antiferromagnetic correlations both along the legs and rungs, whereas figure \ref{fig:ssf}(b) examines $S_z(\pi, 0)$, indicating antiferromagnetic correlations along the legs and ferromagnetic correlations along the rungs. The behavior of the magnetic structure factor along  $(\pi,\pi)$ and  $(\pi,0)$ reveals distinct magnetic arrangements within and outside the Haldane phase. The dome-shaped region in the $(t_d,U)$ plane marking different magnetic arrangements is present for both local (near-neighbor and next-near neighbor spin-spin correlations, Fig.\,\ref{fig:rung_diag_corr}) and long-range (structure factor, Fig.\,\ref{fig:ssf}) magnetic responses. 
Notably, inside the dome, $S_z(\pi,\pi)$ decreases significantly, reflecting suppressed long-range antiferromagnetic order, consistent with the characteristics of the Haldane phase. Outside the dome, larger values indicate more pronounced antiferromagnetic ordering. Thus, these figures solidify our understanding of the magnetic behavior and indicate the presence of two different magnetic arrangements, as illustrated in Figure \ref{fig:mag_ordering}.

\section{Edge correlations}
\label{sec:EC}

In this section, we analyze the edge correlation properties as indicators of the topological character of the Haldane phase \cite{tasaki2020physics}. The presence of correlated edges is a distinctive signature of the Haldane phase. In the spin-1 Heisenberg model, the uniqueness of the ground state, ensured by the Marshall-Lieb-Mattis theorem \cite{lieb1962ordering},\cite{tasaki1990marshall}, leads to nonzero antiferromagnetic edge-edge correlations in systems with open boundary conditions. We capture this feature by computing the spin-spin correlation function between the first rung ($r=0$) and subsequent rungs (up to $r=L-1$) along the ladder, defined as:
\begin{equation}
    \mathcal{O}_{\text{edge}}(r) = \langle T^z_{0} T^z_{r}\rangle
    \label{eqn:rung_edge}
\end{equation}
with $T^z_{r} = S^z_{0,r} + S^z_{1,r}$.

\begin{figure}
    \centering
    \includegraphics[width=\linewidth]{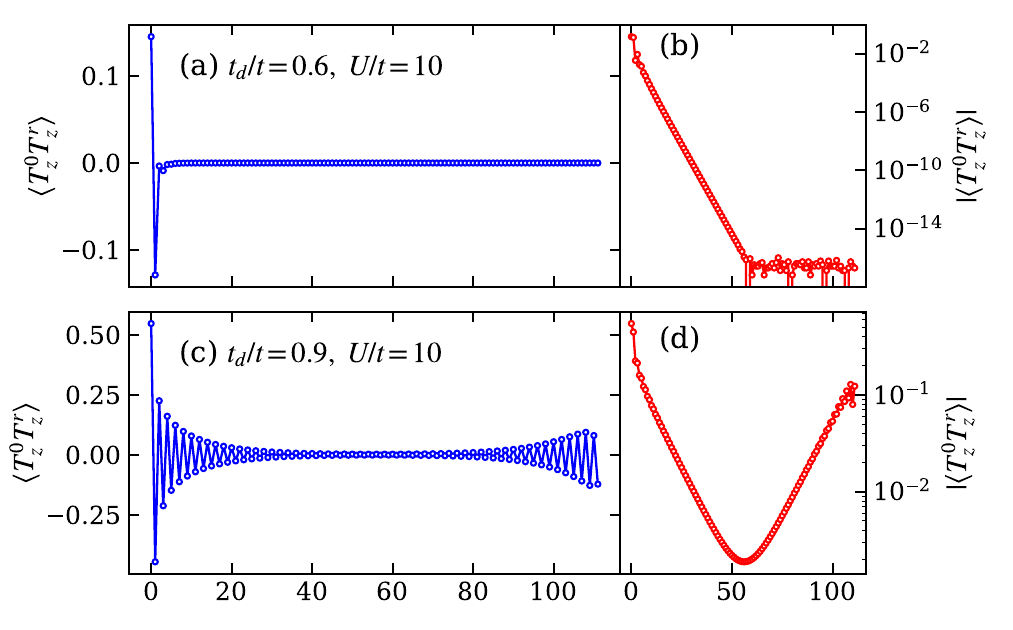}
    \caption{Edge correlations $\langle T^{z}_{0}T^{z}_{r}\rangle$ showing distinct behavior of points outside (panels (a) and (b): $t_d/t = 0.6$, $U/t = 10$) and inside (panels (c) and (d): $t_d/t = 0.9$, $U/t = 10$) the dome for a $L=112$ rung system. Panels (b) and (d) show the modulus of the edge correlation on a logarithmic scale, illustrating the distinct correlation lengths.}
    \label{fig:edge_corr}
\end{figure}

In Figure \ref{fig:edge_corr}, we present the behavior of edge correlations at two distinct parameter points: one outside ($t_d/t = 0.6$, $U/t=10$) and another inside ($t_d/t = 0.9$, $U/t = 10$) the dome-shaped region associated with the Haldane phase. Figures \ref{fig:edge_corr}(a) and \ref{fig:edge_corr}(b) depict correlations within the topologically trivial phase, showing rapid decay towards zero. This behavior aligns with expectations for a globally antiferromagnetic arrangement with no edge-edge correlations. Conversely, figures \ref{fig:edge_corr}(c) and \ref{fig:edge_corr}(d) demonstrate that for a point inside the dome region, edge correlations decay more slowly into the bulk and increase again toward the system's opposite boundary. This slow decay and subsequent rise reflect pronounced edge-edge correlations and indicate the presence of edge states, as expected for the Haldane phase.

The transition between these two phases is captured by the correlation length $\xi_\text{edge}$, defined through the exponential decay of the edge correlation function as:
\begin{equation}
    |\mathcal{O}_\textbf{edge}(r)| \propto e^{-\frac{r}{\xi_\text{edge}}}
    \label{eqn:corr_len}
\end{equation}
capturing the rate with which the correlation function decays.

\begin{figure}[h!]
    \centering
    \includegraphics[width = 0.90\linewidth]{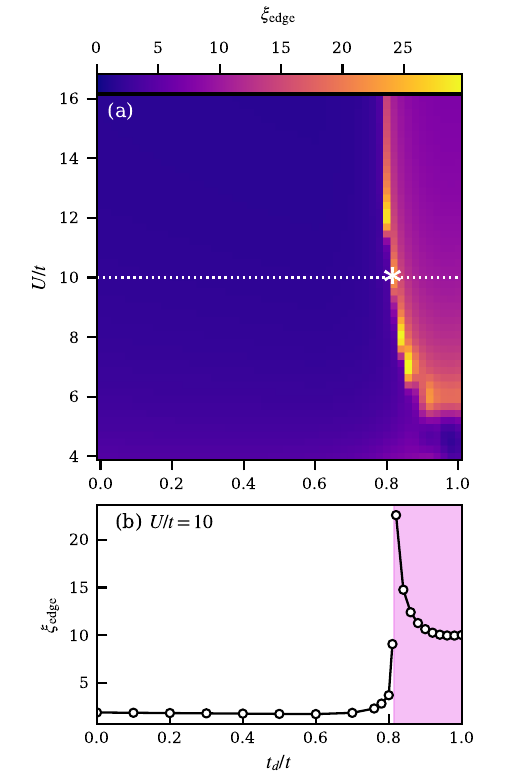}
    \caption{Edge correlation length $\xi_\text{edge}$ as a function of $U/t$ and $t_d/t$ for a $L=112$ rung system. Panel (a) presents a heat map of correlation length across the parameter space, revealing a marked divergence along the phase boundary. The white dotted line corresponds to data shown in panel (b), and the white star marks the entrance to the dome. Panel (b) shows the divergence of $\xi_\text{edge}$ for $U/t = 10$, marking the boundary between the trivial region and the dome region.}
    \label{fig:edgecorr_U10}
\end{figure}

As illustrated in Figure \ref{fig:edgecorr_U10}(a), $\xi_\text{edge}$ exhibits a marked divergence at the transition boundary, separating the trivial phase (outside the dome) from the topologically nontrivial Haldane phase (inside the dome). Figure \ref{fig:edgecorr_U10}(b) explicitly highlights this divergence for a fixed interaction value ($U/t = 10$) as $t_d/t$ varies, confirming that $\xi_\text{edge}$ effectively captures the topological transition.

\section{String order}
\label{sec:SO}

The breakdown of the hidden $\mathbb{Z}_2 \times \mathbb{Z}_2$ symmetry is a hallmark of the Haldane phase. In particular, the presence of hidden antiferromagnetic order is a marker of this breakdown \cite{kennedy1992hidden2}, and can be found through the string order expectation value given by
\begin{equation}
    \label{eqn:string_order}
    \mathcal{O}_\text{string}(r_0,r) = - \left\langle T^z_{r_0} \exp\left(\sum_{l = r_0+1}^{r-1} i\pi T^z_{l} \right)T^z_{r} \right\rangle
\end{equation}
where $T^{z}_r = S^{z}_{0,r} + S^{z}_{1,r}$.
The string order parameter is formally defined as:
\begin{equation}
\mathcal{O}_\text{string} \equiv \lim_{r-r_0 \rightarrow\infty}\lim_{L\rightarrow\infty} \mathcal{O}_\text{string}(r_0,r)
\label{eqn:string_order_parameter}
\end{equation}
The Haldane phase in the spin-1 Heisenberg model is characterized by a nonzero string order parameter, $\mathcal{O}_\text{string}> 0$ \cite{kennedy1992hidden,kennedy1992hidden2}. This topological order does not arise from conventional spontaneous symmetry breaking. Instead, it emerges due to the breakdown of the hidden $\mathbb{Z}_2\times\mathbb{Z}_2$. Specifically, under the Kennedy-Tasaki transformation, given by the unitary operator $U_\text{KT}$ (Eq.\,\ref{eqn:KT}), the Hamiltonian of the spin-1 Heisenberg model,
\begin{equation}
    H_\text{h} = J\sum_r \mathbf{T_r}\cdot\mathbf{T_{r+1}}
\end{equation}
can be mapped to a short-ranged Hamiltonian $H_\text{KT}$ \cite{oshikawa1992hidden}:
\begin{equation}
    H_{\text{KT}} = U_\text{KT} H_\text{h} U_\text{KT}, \quad U_\text{KT} = \prod_{r<l} \exp{i\pi S^z_{r}S^x_{r}}
    \label{eqn:KT}
\end{equation}
which, unlike $H_\text{h}$, does not have full SU(2) symmetry, but is invariant only under rotations of $\pi$ about $x$, $y$ and $z$ directions, such that $H_\text{KT}$ has only the $\mathbb{Z}_2\times\mathbb{Z}_2$ discrete symmetry. The transformed Hamiltonian's ground state
\begin{equation}
    \ket{\psi_\text{KT}} = U_\text{KT}\ket{\psi_\text{h}}
    \label{eqn:trans_gs}
\end{equation}
spontaneously breaks this symmetry exhibiting long-range ferromagnetic order in the $x$ and $z$ direction \cite{tasaki2020physics}, with ferromagnetic order parameter defined by:
\begin{equation}
    \mathcal{O}_\text{ferro}^{\text{KT}} \equiv \lim_{r-r0 \rightarrow\infty}\lim_{L\rightarrow\infty}\bra{\psi_\text{KT}}T^z_{r_0}T^z_{r}\ket{\psi_\text{KT}}
    \label{eqn:ferro}
\end{equation}
We than have $\mathcal{O}_\text{ferro}^\text{KT} > 0$. It can be shown that \cite{tasaki2020physics}:
\begin{equation}
    \mathcal{O}_\text{ferro}^\text{KT} = \mathcal{O}_\text{string}
\end{equation}
establishing a fundamental link between the nonzero string order parameter in the spin-1 Heisenberg ground state and the symmetry breaking of the $\mathbb{Z}_2\times\mathbb{Z}_2$ symmetry in the ground state of the transformed Hamiltonian.

\begin{figure}
    \centering
    \includegraphics[width=\linewidth]{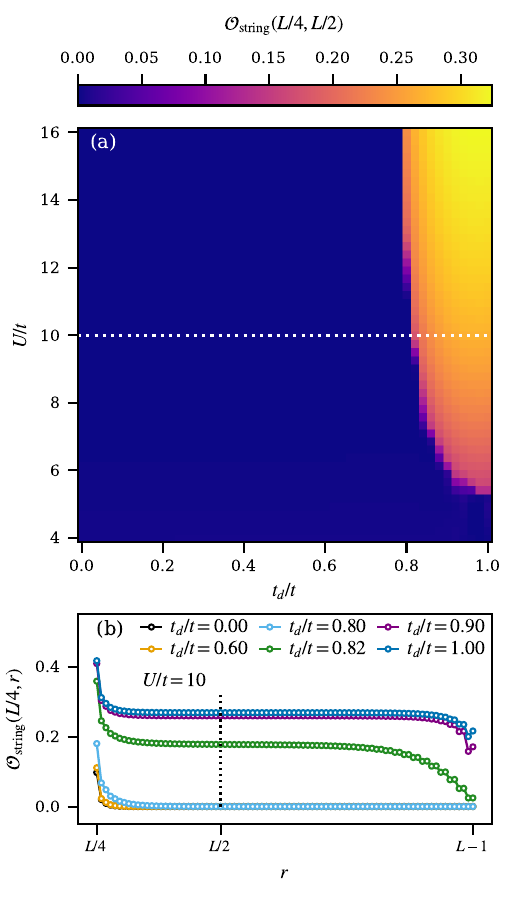}
    \caption{String order expectation value calculated across the parameter space as defined by equation \ref{eqn:string_order} for a $L=112$ rung system. Panel (a) displays a heat map delineating the dome-shaped region associated with the non-trivial Haldane phase. The white horizontal line marks $U/t = 10$, associated with panel (b). Panel (b) shows detailed spatial dependence of the string order parameter at specific points along the $U/t = 10$ line, emphasizing its long-range nature inside the dome.}
    \label{fig:string_order}
\end{figure}

\begin{figure}
    \centering
    \includegraphics[width=\linewidth]{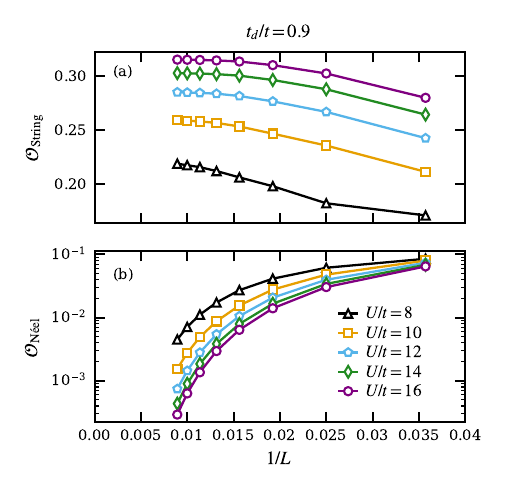}
    \caption{Finite-size scaling analysis of order parameters within the Haldane region for $t_d/t = 0.9$. Panel (a) demonstrates that the string order parameter stabilizes to a nonzero value in the thermodynamic limit. Panel (b) shows the Néel order parameter scaling towards zero with increasing system size, confirming the absence of conventional magnetic order within the Haldane phase.}
    \label{fig:soneel}
\end{figure}

Figure \ref{fig:string_order}(a) shows the behavior of the string order expectation value on the ($t_d/t$,$U/t$) plane. Positive values are distinctly observed within the dome-shaped region associated with the Haldane phase. Figure \ref{fig:string_order}(b) further details the spatial behavior of this parameter at specific points, highlighting the long-range string order within the topological region.
Additionally, Figure \ref{fig:soneel} provides a finite size analysis of the string and Néel order parameters to verify robustness in the thermodynamic limit. Figure \ref{fig:soneel}(a) shows the increase and stabilization of the string order parameter with increasing system size for $t_d/t = 0.9$, pointing to the robustness of the hidden order. In contrast, Figure \ref{fig:soneel}(b) demonstrates the Néel order parameter diminishing towards zero with increasing system size, confirming that the observed order within the dome is exclusively topological and not trivial (no spontaneous symmetry breaking) \cite{kennedy1992hidden}. 

It is worth pointing out that the phase transition from the topologically trivial phase to the Haldane phase is being signaled by all quantities calculated so far. In Figure \ref{fig:rung_diag_corr}(c) the transition from positive to negative spin-spin diagonal correlation for $U/t = 10$ occurs between $t_d/t = 0.8$ and $t_d/t = 0.82$, the same interval where the divergence in the correlation length $\xi_\text{edge}$ occurs [Figure \ref{fig:edgecorr_U10}(b)] and where the string order becomes long-ranged [Figure \ref{fig:string_order}(b)]. The points in the phase diagram in Figure \ref{fig:phase_diagram} were obtained by observing when the string order becomes long-ranged for different values of $U/t$, demarcating the topological dome region.

\section{Spin Gap}
\label{sec:SG}

\begin{figure}
    \centering
    \includegraphics[width=\linewidth]{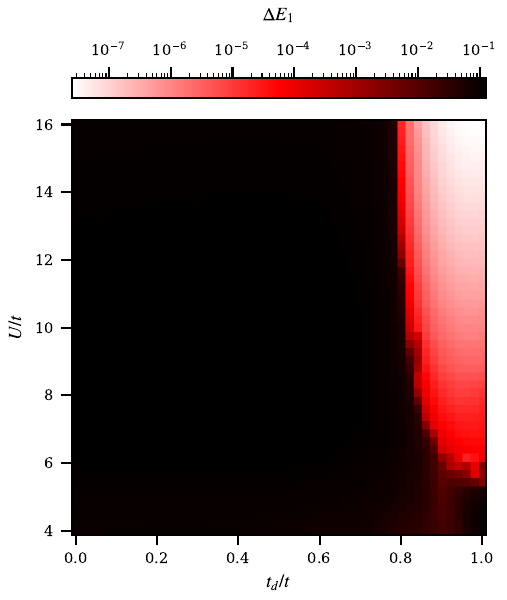}
    \caption{Spin gap calculated according to equation \ref{eqn:spin_gap1} as a function of $t_d/t$ and $U/t$ for a $L=112$ rung system. The logarithmic color scale emphasizes the suppression of the spin gap in the parameter region corresponding to the Haldane dome.}
    \label{fig:spin_gap1}
\end{figure}

\begin{figure}
    \centering
    \includegraphics[width=\linewidth]{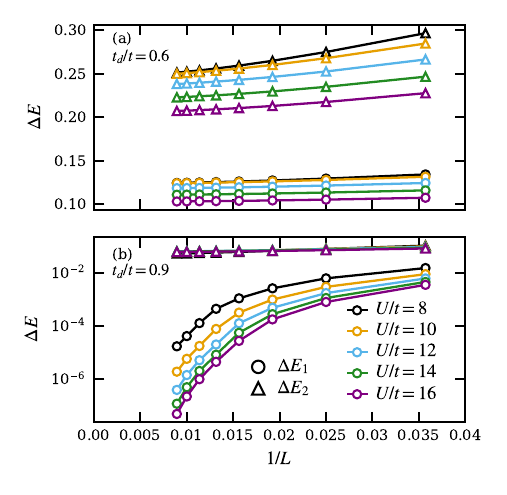}
    \caption{Finite-size scaling analysis of the first ($\Delta E_1$) and second ($\Delta E_2$) spin gaps for representative points outside (a) $t_d/t = 0.6$ and inside (b), $t_d/t = 0.9$ the Haldane dome. Within the Haldane phase, $\Delta E_1$ decreases exponentially for large $L$, while $\Delta E_2$ remains finite, reflecting the Haldane gap.}
    \label{fig:spin_gap_scale}
\end{figure}

In the Haldane phase, there is the presence of a triplet of excited states, known as the Kennedy triplet, with an exponentially decaying gap from the ground state with increasing system size \cite{yamamoto1997low}. In these excited states with $S^\text{tot} = 1$, the additional spin excitation is exponentially localized at the edges, further emphasizing its topological nature \cite{qin1995edge}. As the thermodynamic limit is approached, the edge-edge correlation (equation \ref{eqn:rung_edge}) vanishes, and the $S^\text{tot} = 0$ to $S^\text{tot} = 1$ transition is accompanied by gapless edge excitations, making the ground state and the Kennedy triplet degenerate. The spin gap is defined as:
\begin{equation}
    \label{eqn:spin_gap1}
    \Delta E_k = E(N_{\uparrow} = L+k,N_{\downarrow} = L-k) - E(N_{\uparrow} = L,N_{\downarrow} = L) 
\end{equation}
Figure \ref{fig:spin_gap1} presents the spin gap  $\Delta E_1$ (taken from equation \ref{eqn:spin_gap1} with $k=1$) in the  ($t_d/t$,$U/t$) plane, showing the transition into the topological regime. Points located inside the dome-shaped region associated with the Haldane phase exhibit a vanishing, yet finite $\Delta E_1$, indicative of the closing of the lowest spin gap.
In contrast to the $\Delta E_1$ case, the presence of a nonzero energy gap (the Haldane gap $\Delta E_2$) in the thermodynamic limit for a system with unbroken continuous symmetry is one of the key manifestations of the Haldane phase \cite{tasaki2020physics}. Figure \ref{fig:spin_gap_scale} further illustrates the scaling behavior of the spin gap. Specifically, Figure \ref{fig:spin_gap_scale}(a) shows that in the topologically trivial phase ($t_d/t = 0.6$), both spin gaps ($\Delta E_1$ and $\Delta E_2$) remain finite as the system size increases, indicating a trivial gapped state. Conversely, Figure \ref{fig:spin_gap_scale}(b) illustrates the behavior in the topological phase ($t_d/t = 0.9$), where the first spin gap ($\Delta E_1$) decreases exponentially toward zero with increasing system size, consistent with the presence of gapless edge excitations. Meanwhile, the second spin gap ($\Delta E_2$) remains finite, reflecting the existence of the Haldane gap.

\begin{figure}
    \centering
    \includegraphics[width=\linewidth]{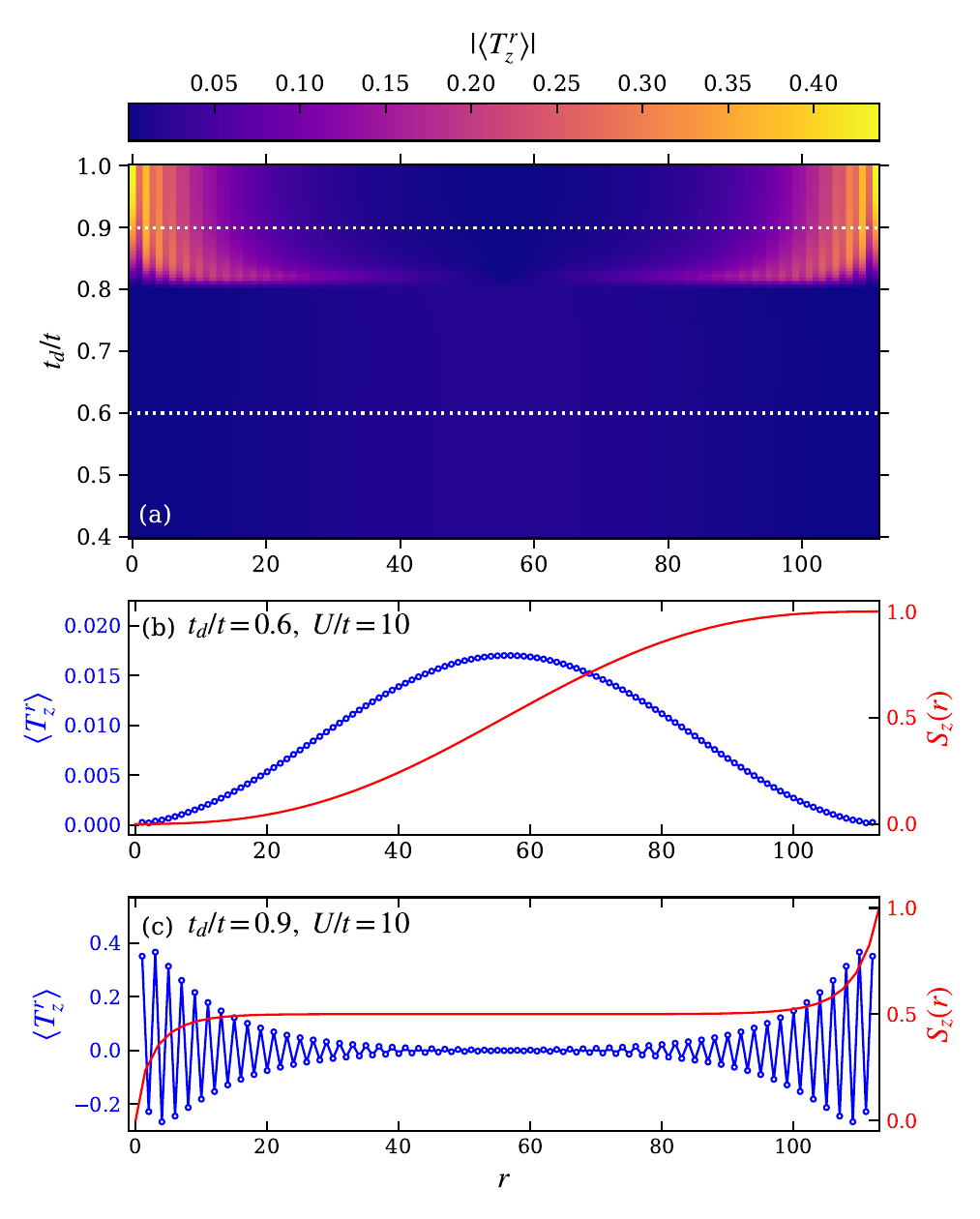}
    \caption{Spatial distribution of spin excess along the ladder, highlighting edge localization as a marker of topological order for a $L = 112$ rung system. Panel (a) shows a heatmap of the spin distribution as a function of $t_d/t$ at fixed $U/t = 10$. The dotted lines highlight $t_d/t = 0.6$ (topologically trivial phase) and $t_d/t = 0.9$ (Haldane phase), shown in panels (b) and (c) respectively where we compare the spatial profiles of the excess spin outside and inside the dome,  showing exponential localization of spin-1/2 edge states within the Haldane phase.}
    \label{fig:spin1_edges}
\end{figure}

Figure \ref{fig:spin1_edges} complements this analysis by depicting the rung spin distribution for the excited states, underlining how the local magnetic moments is distribute across the system. Outside the dome region, the systems spin is concentrated in the bulk of the ladder, indicative of a trivial phase [Figure \ref{fig:spin1_edges}(b)]. In contrast, inside the dome, the spin distribution becomes exponentially localized at the ladder edges [Figure \ref{fig:spin1_edges}(c)] (see red curve given by equation \ref{eqn:total_spin} below), indicating the formation of localized spin-1/2 edge states.
\begin{equation}
    S_z(r) = \sum_{l=0}^{r-1}\langle T^z_l + T^z_{l+1}\rangle
    \label{eqn:total_spin}
\end{equation}
This localization further proves the topologically nontrivial nature of the phase inside the dome-shaped region. Figure \ref{fig:spin1_edges}(a) emphasizes this transition visually, showing how the spin distribution across rungs changes distinctly as the system crosses into the dome region.

In agreement with the previous discussion, in Figure \ref{fig:spin1_edges}(a), the spin distribution for $U/t = 10$ shows that the presence of edge states starts to occur between $t_d/t = 0.8$ and $t_d/t = 0.82$.  

\section{Entanglement Spectrum}
\label{sec:ES}

Although the presence of edge states, Kennedy triplet, and a nonzero string order strongly suggest the Haldane phase, prior studies indicate that charge fluctuations might allow an adiabatic connection between trivial and topologically nontrivial states \cite{anfuso2007string,moudgalya2015fragility}. In this context, the entanglement spectrum is a decisive indicator of topology \cite{pollmann2010entanglement}.

We divide our system into two partitions: a left block $\mathcal{L}$ and a right block $\mathcal{R}$, such that the system's Hilbert space is the tensor product: $\mathcal{H} = \mathcal{H_L}\bigotimes \mathcal{H_R}$, where $\mathcal{H_{L(R)}}$ is the Hilbert space of the sites on the left(right) partition. Using the Schmidt decomposition, the ground state can be expressed as: 
\begin{equation}
    \label{eqn:schmidt_decomposition}
    \ket{\psi} = \sum_{i} \lambda_i \ket{\lambda^\mathcal{L}_i}\otimes\ket{\lambda^\mathcal{R}_i}
\end{equation}
where the $\{\ket{\lambda^{\mathcal{L(R)}}_{i}}\}$ forms an orthonormal basis for $\mathcal{H_{L(R)}}$ and the Schmidt values $\{\lambda_{i}\}$ constitute the entanglement spectrum.

It has been established that the Haldane phase features a characteristic degeneracy in the entanglement spectrum directly associated with inversion, time-reversal, or $\mathbb{Z}_2\times \mathbb{Z}_2$ symmetry. \cite{pollmann2012symmetry,pollmann2010entanglement}. Specifically, all Schmidt values appear with even degeneracy in the Haldane phase, reflecting the underlying topological order.
The implications of this degeneracy are significant for the entanglement entropy:
\begin{equation}
    \label{eqn:von_neumann_entropy}
    S_{\mathcal{L}|\mathcal{R}} = -\sum_i \lambda_i^2\log{\lambda_i^2}
\end{equation}
The minimum entropy for an evenly degenerate spectrum occurs when exactly two Schmidt values are non-zero and equal, in particular $\lambda_1 = \lambda_2 = 1/\sqrt{2}$, leading to:
\begin{equation}
    \label{eqn:minimum_entanglement_entropy}
    \text{min}(S_{\mathcal{L}|\mathcal{R}}) = \log{2}
\end{equation}



This property underlies the impossibility of adiabatically connecting a nontrivial topological state to a trivial product state. To quantitatively assess the degeneracy, we calculate the largest gap in the Schmidt spectrum \cite{jazdzewska2023transition}, defined as: 
\begin{equation}
    \label{eqn:largest_gap}
    \Delta_\lambda = \text{max}(\delta_1,\delta_2,\cdots) 
\end{equation}
where $\delta_n = \text{min}(\ln\lambda_n - \ln\lambda_{n-1}, \ln\lambda_{n+1} - \ln\lambda_{n})$. A fully degenerate spectrum yields $\Delta_\lambda = 0$ whereas the presence of non-degenerate Schmidt values results in a finite $\Delta_\lambda$.

\begin{figure}
    \centering
    \includegraphics[width=\linewidth]{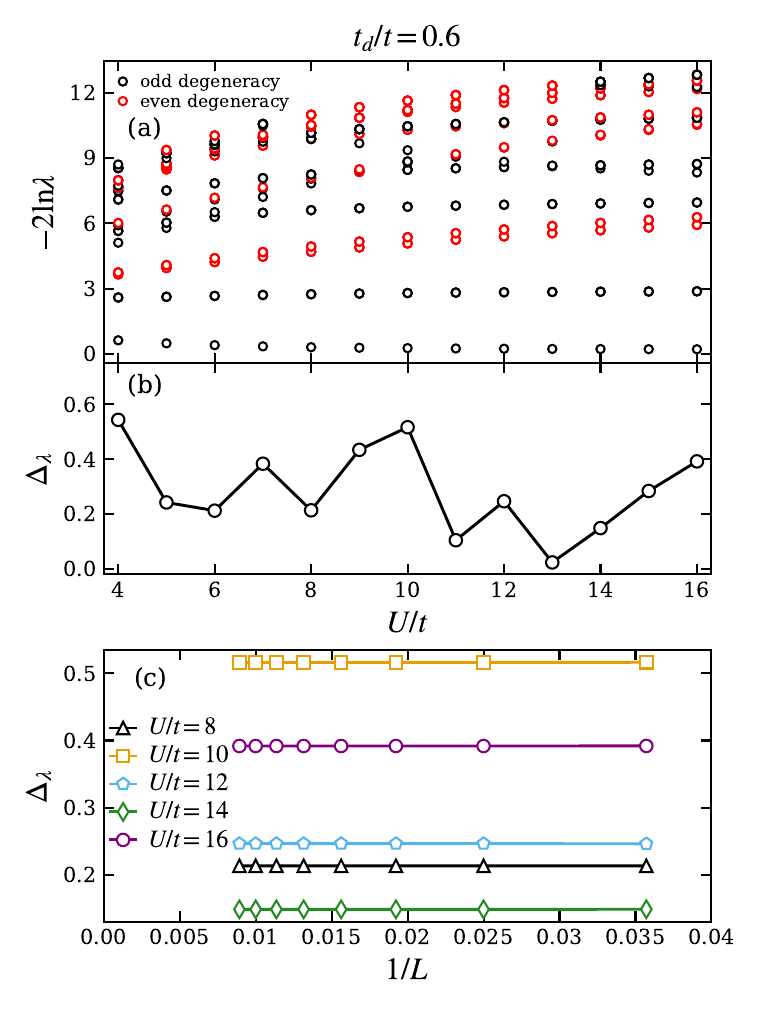}
    \caption{Entanglement spectrum analysis at $t_d/t = 0.6$ (topolologically trivial phase) for a $L = 112$ rung system partitioned in half. Panel (a) shows the Schmidt eigenvalues with no trend towards even degeneracy, panel (b) the largest entanglement gap remaining finite, and panel (c) the lack of size dependence on the largest gap.}
    \label{fig:ent_spec_0.6}
\end{figure}

\begin{figure}
    \centering
    \includegraphics[width=\linewidth]{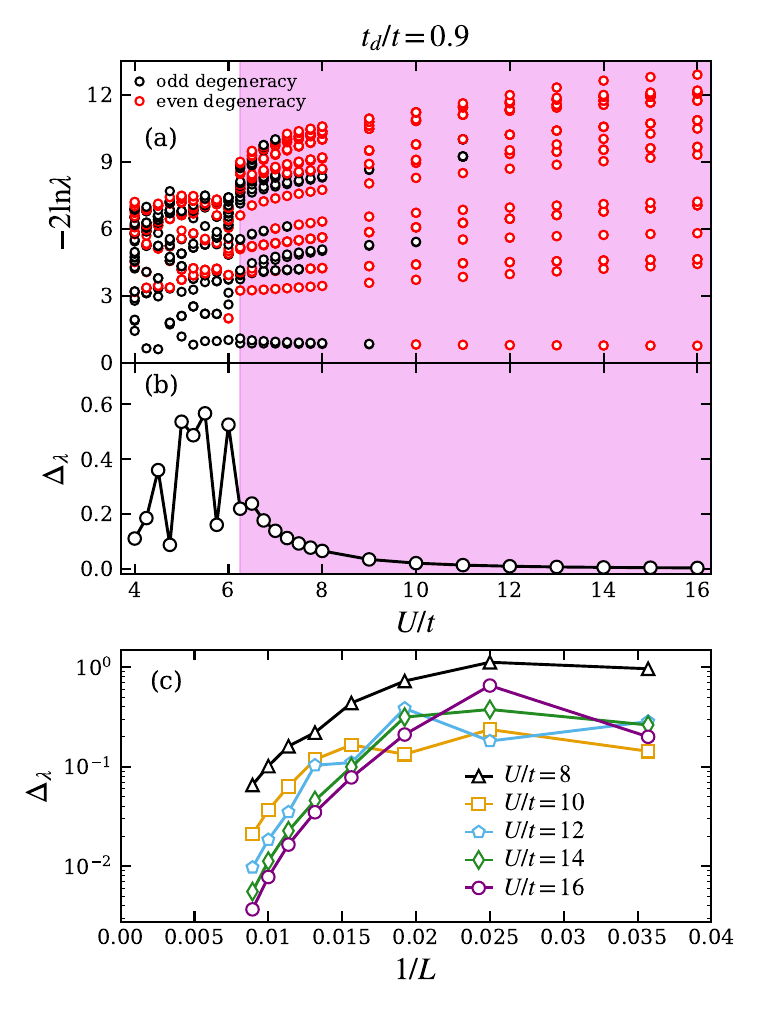}
    \caption{Entanglement spectrum and largest entanglement gap at $t_d/t = 0.9$ (within the Haldane phase) for a $L = 112$ rung system partitioned in half. Panel (a) reveals the Schmidt values even degeneracy in the topological phase, (b) the largest gap approaching zero with increasing $U/t$, and panel (c) demonstrates the largest gap decreasing as the system size grows, strongly indicating the presence of a symmetry-protected topological phase. The violet region in panels (a) and (b) correspond to the $U/t$ values where the system for $t_d/t = 0.9$ is inside the dome-shaped region that marks the topological phase.}
    \label{fig:ent_spec_0.9}
\end{figure}

Figures \ref{fig:ent_spec_0.6} and \ref{fig:ent_spec_0.9} examine the entanglement spectrum and its largest gap for points outside and inside the dome shaped region that marks the topological phase border, respectively. Figure \ref{fig:ent_spec_0.6}(a) presents the entanglement spectrum for $t_d/t = 0.6$, showing no tendency towards an evenly degenerate structure as $U/t$ increases, indicating of a trivial phase. Correspondingly, Figure \ref{fig:ent_spec_0.6}(b) shows that the largest gap, $\Delta_\lambda$, remains finite and exhibits no significant scaling with system size [Figure \ref{fig:ent_spec_0.6}(c)], further confirming the trivial nature of this region in the thermodynamic limit.


In contrast, Figure \ref{fig:ent_spec_0.9}(a) illustrates the entanglement spectrum at $t_d/t = 0.9$, showing an evolution towards even degeneracy within the topological phase (violet region) as $U/t$ increases. Figure \ref{fig:ent_spec_0.9}(b) further reinforces this observation, demonstrating a distinct reduction in the largest gap, $\Delta_\lambda$, which approaches zero with increasing $U/t$. Additionally, Figure \ref{fig:ent_spec_0.9}(c) shows that $\Delta_\lambda$ systematically decreases as the system size grows, supporting the existence of a robust and nontrivial symmetry-protected topological phase.

\section{Charge Gap}
\label{sec:CG}

\begin{figure}
    \centering
    \includegraphics[width=\linewidth]{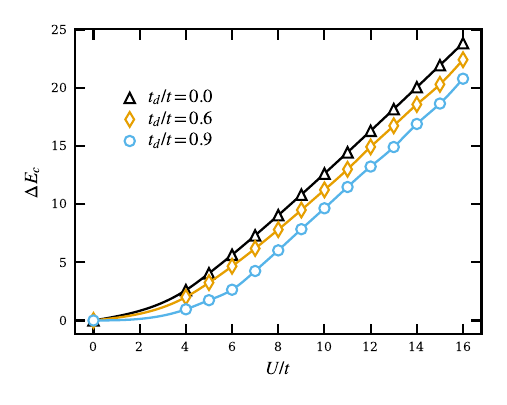}
    \caption{Charge gap $\Delta E_c$ as a function of the on-site interaction $U$ for several values of diagonal hopping ($t_d/t$) for a $L=112$ rung system. Lines are guides to the eye.}
    \label{fig:cg}
\end{figure}

In the limit of zero diagonal hopping ($t_d/t = 0$), the system is a Mott insulator [36]. To characterize the insulating behavior as a function of the on-site interaction U for different $t_d/t$ values, we define the charge gap $\Delta E_c$ as
\begin{equation}
\begin{gathered}
    \Delta E_c = E(L+1,L+1) + E(L-1,L-1) \\- 2E(L,L)
\end{gathered}
\label{eqn:charge-gap}
\end{equation}
where $E(N_\uparrow,N_\downarrow)$ denotes the ground-state energy for a configuration with $N_\uparrow$ spin-up and $N_\downarrow$ spin-down fermions. In the non-interacting limit ($U = 0$) and for $t = t_\perp$, the band structure (not shown) doesn't display a gap, and the system remains metallic for all $0<t_d/t<1$. However, as U increases, a charge gap opens and, in the strong-coupling regime, $\Delta E_c$ exhibits a linear dependence on U, reflecting the formation of the Mott-Hubbard gap. This linear scaling, as shown in Figure \ref{fig:cg}, underscores the role of electron-electron interactions in driving the system from a metallic to a Mott insulating state.

\section{Discussion}
\label{sec:D}

In this study, we have investigated the topological phase transition in the fermionic two-leg Hubbard model, emphasizing the critical role of diagonal hopping ($t_d/t$) and strong electron-electron interactions ($U/t$). By employing the DMRG method, we identified a distinct dome-shaped region in the ($t_d/t, U/t)$ plane where the topologically nontrivial Haldane phase emerges.

Through analyses of spin-spin correlations, edge correlations, string order parameters, spin gaps, and entanglement spectra, we established clear signatures that characterize the topological nature of the Haldane phase. Notably, we observed a significant change in diagonal spin-spin correlations, transitioning from positive to negative values, which signals the predominance of effective rung-triplet states.

The presence of nontrivial string order and robust edge correlations further confirmed the hidden antiferromagnetic order and of localized spin-1/2 edge states, characteristic of the Haldane phase. Additionally, the scaling analysis of the spin gap and entanglement spectrum demonstrate the presence of the Haldane gap, gapless edge excitations, and an evenly degenerate entanglement spectra in the thermodynamic limit, reinforcing the system's topological robustness.

As for the trivial phase, the charge gap indicates it is a Mott insulator, as the charge gap grows linearly with $U$ in the strong-coupling regime, following the $t_d/t = 0$ limit case.

Our findings underline the importance of both frustration (introduced by diagonal hopping) and strong interactions in driving the system toward nontrivial symmetry-protected topological phases. The presence of a topological phase for large $U$ and $t_d/t=1$ was expected from the mapping on the spin-1 Heisenberg chain. Interestingly, we found the Haldane phase extends beyond this particular point, giving rise to an extended region in the $(t_d/t,U/t)$ plane. Recent experiments with ultra-cold fermionic atoms have studied Hubbard two-leg ladders  \cite{sompet2022realizing} in a similar setting as the one proposed here. 
Our study provides significant insights into the emergence and characterization of topological phases in ladder systems, including the location of the phase boundary between a trivial and a topological phase, providing guidance for future experimental exploration in strongly correlated ladder materials and quantum simulation platforms.

\section*{ACKNOWLEDGMENTS}

The authors are grateful to 
Coordena\c c\~ao de Aperfei\c coamento de Pessoal de Ensino Superior (CAPES), and Instituto Nacional de Ci\^encia e Tecnologia de Informa\c c\~ao Qu\^antica (INCT-IQ) for funding this project. 
We also gratefully acknowledge support from Funda\c c\~ao Carlos Chagas de Apoio \`a Pesquisa (FAPERJ), through the grants  
E-26/200.959/2022  (T.P.), 
and  E-26/210.100/2023  (T.P.); 
and from Conselho Nacional de Desenvolvimento Cient\'\i fico e Tecnol\'ogico through the grants 
308335/2019-8 (T.P.), 
403130/2021-2 (T.P.), 
and 442072/2023-6 (T.P.). 

\label{Bibliography}
\bibliography{Bibliography}  
\end{document}